\documentclass[prd, preprint, 12pt]{revtex4-1}

\usepackage{amsmath}
\usepackage{amssymb}
\usepackage{setspace}
\usepackage{graphicx}
\usepackage{bbm}
\usepackage{float}
\usepackage{hyperref}
\usepackage{footnote}

\begin{document}
 
 %

\begin{center}
 {  \large {\bf Quantum theory and the structure of space-time}}



{\bf Tejinder Singh}

\smallskip

{\it Tata Institute of Fundamental Research,}
{\it Homi Bhabha Road, Mumbai 400005, India}

\end{center}

\setstretch{1.24}


\centerline{\bf ABSTRACT}

\noindent We argue that space and space-time emerge as a consequence of dynamical collapse of the wave function of macroscopic objects. Locality and separability are properties of our approximate, emergent  universe. At the fundamental level, space-time is non-commutative, and dynamics is non-local and non-separable.



\noindent 




\section{Space-time is not the perfect arena for quantum theory}
\noindent Space-time is absolute, in conventional classical physics. Its geometry is determined dynamically by the distribution of classical objects in the universe. However, the underlying space-time manifold is an absolute given, providing the arena in which material bodies and fields exist. The same classical space-time arena is carried over to quantum mechanics and to quantum field theory, and it works beautifully there too. Well, almost! But not quite. In this essay we propose the thesis that the troubles of quantum theory arise because of the illegal carry-over of this classical arena. By troubles we mean the quantum measurement problem, the spooky action at a distance and the peculiar nature of quantum non-locality, the so-called problem of time in quantum theory, the extreme dependence of the theory on its classical limit, and the physical meaning of the wave function \cite{Weinberg:2017}. We elaborate on these in the next section. Then, in Section III, we propose that the correct arena for quantum theory is a non-commutative space-time: here there are no troubles. Classical space-time emerges as an approximation, as a consequence of dynamical collapse of the wave function. When quantum theory is viewed on the
 non-commutative space-time, there is no notion of distance, no spooky action at a distance, no mysterious probabilities.
 

\section{Quantum theory is the truth but not the whole truth }
\noindent Let us accept as given, the universe with its terrestrial physics laboratories, planets, stars, galaxies, and the classical absolute background of space and time. When we treat the microscopic world according to the laws of quantum mechanics, against this given background, we have a theory which agrees marvellously with every experiment that has been done so far to test it. Even in the macroscopic world, the theory exhibits many collective phenomena such as superconductivity, superfluidity, and Bose-Einstein condensation. However, as is well-known, despite its great success, there are some features of the theory that many physicists find tentative, and/or unsatisfactory. We recall these briefly, in the next few paragraphs.  

\medskip

$\bullet$ {\textbf {\textit {The theory has not yet been experimentally tested in all parts of the  parameter space}}}: The principle of linear superposition is  a central tenet of quantum theory, and is famously shown to hold in the double slit experiment with photons and with electrons. It has been experimentally shown to hold also for heavier particles such as neutrons, and atoms, and molecules. But what about heavier composite systems such as macromolecules or even larger objects, say nano-particles, or silicon clusters? Technology poses great challenge to matter-wave interferometry experiments, or any other kind of experiments, which aim to test the superposition of position states for mesoscopic objects. The largest macromolecule for which superposition in an interference experiment has been shown to hold, has a mass
of about 10,000 atomic mass units \cite{Arndt:2014}. On the other hand, in the  macroscopic classical world, position superposition does not seem to hold: a chair is never seen here and there at the same time. The smallest objects for which such classical behaviour is known to hold for sure, have a mass of about a microgram, which is about $10^{18}$ atomic mass units. Now it could well be that this absence of macroscopic superposition is caused by environmental decoherence, in a many-worlds scenario. Or it could be explained by Bohmian mechanics or some other interpretation / reformulation of quantum theory. But how can we be sure that this is not a result of new physics: the possibility that linear superposition is not an exact but an approximate principle? With superposition lasting for enormously long times for small particles such as electrons and atoms, but becoming progressively short-lived for heavier objects, until it becomes extremely short-lived for macroscopic objects such as chairs and tables. The only way to settle whether or not there is new physics is to do experiments in a decoherence free environment. These experiments are extremely challenging, and are being pursued in many laboratories that exploit frontline technology \cite{RMP:2012}. Between 10,000 a.m.u. and $10^{18}$ a.m.u. there is an experimentally untested desert spanning fourteen orders of magnitude, where there is a question mark on the validity of the superposition principle. [Note that collective internal states such as in superconductors and BECs do not qualify as experimental tests of superposition in the present context, because these `one-particle type' states are not the Schr\"{o}dinger cat states one is looking to create].

\medskip

$\bullet$ {\textbf {\textit{The quantum measurement problem}}}: When a quantum system, which is in a superposition of eigenstates of an observable, interacts with a classical measuring apparatus, why does it end up in being just one of those eigenstates? What causes this breakdown of superposition and the so-called collapse of the wave function, in apparent violation of the Schr\"{o}dinger equation? And what systems qualify to be called a classical measuring apparatus? Quantitatively speaking, how large must a quantum system be, before it can be called classical? And why is the collapse random, even though the Schr\"{o}dinger equation is deterministic? Where do probabilities come from, in a deterministic theory, where initial conditions are perfectly well-defined? Standard quantum theory has no answers. Do these unanswered questions call for new physics? Maybe! \cite{Gisin:2017}.

\medskip

$\bullet$ {\textbf {\textit {The peculiar nature of quantum non-locality}}}: Consider a pair of quantum particles A and B, in an entangled state, which start off close by, and fly off in opposite directions to get very, very far from each other. Perhaps so far that they travel for billions of light years while remaining in an entangled state. If a measurement is made say on A, causing its state to collapse, it instantly and acausally influences the state of B, as if A and B were together a physical structure like a rigid body. This is the non-local quantum correlation, the infamous spooky action at a distance \cite{Musser:2015}. And it has been confirmed by 
experiments \cite{loophole2015,loophole22015,loophole32017}. The measurement on A influences the state of B, even though B is outside the light-cone of A [space-like separation]. Of course the measurement on A cannot be used to transfer information to B faster than light (no superluminal signalling) \cite{Ghirardi:80}. This is the microcausality condition: quantum commutators vanish outside the light-cone. Is non-locality a problem? For many physicists, it is not a problem, because special relativity is not violated. This is how quantum physics is, they say. For other physicists, which includes the present author, quantum non-locality 
is one of the greatest mysteries of physics. It challenges cherished and long-held notions of locality and separability. How does B `know' that a measurement has been made on A? Undoubtedly, the collapse aspect of quantum theory is oblivious to physical distance \cite{Bell:86}.

\medskip

$\bullet${\textbf {\textit {The problem of time in quantum theory}}}: The time that appears in quantum theory is part of a classical space-time, whose geometry is determined by macroscopic classical bodies, according to the laws of general relativity. But these classical bodies are a limiting case of quantum theory. In their absence [for instance in the very early universe, soon after the big bang] there would be no classical space-time geometry. If there are only quantum matter fields in the entire universe, the gravitational field they produce would also possess quantum  fluctuations. These fluctuations in turn destroy the underlying 
space-time manifold, rendering it physically meaningless. This is the essence of the Einstein hole argument: a classical metric field must overlie the space-time manifold, in order to allow a physical interpretation of the point structure of the manifold \cite{Carlip2001}. How then does one describe quantum dynamics of quantum objects if the universe is not dominated by macroscopic objects, and does not possess a classical space-time manifold? As a matter of principle one ought to be able to describe the dynamics under such circumstances. It therefore follows that there must exist an equivalent reformulation of quantum theory which does not depend on classical time \cite{Singh:2006}.

\medskip

$\bullet$ {\textbf{\textit {The extreme dependence of the theory on its classical limit}}}: Classical mechanics is a limiting case of quantum mechanics. Yet, in order to arrive at the canonical quantum theory, one must first know the classical theory. One must know the classical configuration and momentum degrees of freedom, its Lagrangian, its Hamiltonian, its Poisson brackets. Then, employing an ad hoc [albeit highly successful] recipe, the Poisson brackets are replaced by the canonical commutation relations of quantum theory. This dependence of the quantum theory on its classical limit is unsatisfactory. It is as if we have a set of quantum recipes; rather than a theory constructed from first principles. There ought to exist a construction / derivation of quantum theory which does not depend on its own classical limit.

\medskip

$\bullet${\textbf{\textit{The physical meaning of the wave function}}}: In order to illustrate the difficulty, let us consider the double slit experiment with electrons. An electron, supposedly behaving like a localised particle before it leaves the electron gun, spreads like a wave after leaving the gun. The wave passes through both the slits, the two secondary wavelets interfere with each other, until the electron reaches the screen and randomly collapses to a point on the screen. But wait a moment! What is this wave a wave of? In order to describe the interference pattern, one must consider the `wave' to be the wave function (i.e. the so-called probability amplitude) and superpose the complex wave functions representing the two wavelets. But how could the complex wave function live on real physical space, and pass through the slits, and behave as if it were a wave? The wave function  in fact lives in Hilbert space, not in physical space. To save the phenomenon, the wave is sometimes said to be a probability wave. But of course we know that probabilities to do not interfere, probability amplitudes do. And if it is a probability wave, what physical meaning can we ascribe to the wave function? This bizarre state of affairs led Feynman to famously assert that nobody understands quantum mechanics. We have an extraordinarily successful theory, but we have a rather poor idea what actually is going on out there.

To understand what is going on, we take the holistic view that all these problems described above must have a common resolution. Quantum theory is not the whole truth. It is an approximation to a deeper theory.  A resolution to its problems does exist, and when we push the proposed resolution to its logical conclusion, we are confronted with a new conception of space and time. Space and time, as we know them, are emergent, and result from dynamical collapse of the wave function. Fundamentally, space-time is non-commutative, and its geometry is a non-commutative geometry.

\section{The whole truth is that quantum theory is approximate}
\medskip
$\bullet$ {\textbf{\textit {A possible resolution of the quantum measurement problem}}}: There have of course been various proposed solutions for the measurement problem, which result from a reinterpretation or mathematical reformulation of quantum theory. These include the many-worlds interpretation, Bohmian mechanics, consistent histories, amongst others. Since they make the same experimental predictions as quantum theory, it really becomes a matter of taste as to which one is preferred over the other \cite{RMP:2012}.

On the other hand, consider the possibility that quantum theory actually breaks down in the domain of measurement, and is not the right theory to describe the interaction of a quantum system with a measuring apparatus. Maybe there is a more general universal dynamics, which includes a stochastic element responsible for the randomness, and which reduces to quantum mechanics for microscopic systems, and to classical mechanics for macroscopic systems. In the micro- limit, the stochastic aspect is extremely negligible, and in the macro- limit it is so significant that the dynamics looks like classical dynamics.

Any attempt to generalise the Schr\"{o}dinger equation must obey severe constraints, if it has to solve the measurement problem, without contradicting the outstanding experimental success of the Schr\"{o}dinger equation. The new dynamics must be non-linear so that it causes breakdown of superposition during a measurement. The non-linear aspect must be random in nature, so that outcomes of measurement are random, and also because deterministic non-linearities lead to superluminal signalling, which we wish to avoid. The new dynamics has to be non-unitary, in order that the system is driven to one outcome; yet the non-unitary evolution must be norm-preserving, so that probability is conserved. The universal dynamics must also possess an amplification mechanism, so that the non-linearity is negligible for small masses, and becomes progressively more significant for larger masses. 

Remarkably enough, such a dynamics does exist, and was proposed many years ago by Ghirardi, Rimini, Weber and Pearle \cite{Pearle:76,Ghirardi2:90,Ghirardi:86,Pearle:89,Bassi:03}. It is known as Continuous Spontaneous Localisation [CSL]. It proposes that the wave function of every elementary particle (let us assume it to be the nucleon) undergoes a spontaneous collapse
at the rate $\lambda$, and gets localised in physical space to a region of size $r_C$. Between every  two collapses, the wave function follows the usual Schr\"{o}dinger evolution.   $\lambda$ and $r_C$ are two new constants of nature, which in the CSL model are assumed to take the values
\begin{equation}
\lambda \sim 10^{-17} {\rm s^{-1}} ; \qquad r_C \sim 10^{-5} {\rm cm} 
\end{equation}
The wave function of a composite object consisting of $N$ nucleons collapses at the rate $N\lambda$: this is the desired amplification mechanism. The CSL effect is mathematically described by a stochastic non-linear modification of the Schr\"{o}dinger equation which satisfies the constraints mentioned above. CSL 
provides a solution of the measurement problem, explains the classical nature of macroscopic objects, and explains the Born probability rule, by converting linear superposition into an approximate principle. The theory makes predictions for experiments, which differ from the predictions of quantum theory, in the currently untested mesoscopic domain. Ongoing laboratory experiments are at present in an exciting and crucial phase, and could verify or rule out CSL in the next few years \cite{RMP:2012,Carlesso:2016,Vinante:2016a,Bahrami:2014a,Bera:2015}.

A serious limitation of the CSL model is that it is non-relativistic, and dedicated efforts to make a relativistic version face serious difficulties. And perhaps for good reason. We will argue below that collapse and relativity are incompatible, and the non-locality induced by collapse can be meaningfully understood only in a non-commutative space-time obeying a generalised Lorentz invariance. 

\medskip

$\bullet$ {\textbf{\textit {Collapse models are phenomenological: what is their fundamental origin?}}}: 
Another limitation of dynamical collapse models, of which CSL is the most advanced, is that they are purely phenomenological, having been designed for the express purpose of solving the quantum measurement problem. A more fundamental theory would derive the CSL model from some underlying first principles. A beautiful attempt in this direction has been made in the theory of Trace Dynamics [TD] developed by Adler and collaborators \cite{Adler:94,Adler-Millard:1996,Adler:04}. TD is a Poincar\'e invariant classical dynamics theory in which the fundamental canonical degrees of freedom are matrices whose elements are complex valued Grassmann variables. One constructs the Lagrangian and Hamiltonian dynamics of this classical theory in the usual way, and the configuration and momentum variables $q_i$ and $p_i$, which are matrices (equivalently operators), all obey arbitrary  commutation relations with each other. Nonetheless, as a result of a global unitary invariance, the theory possesses the remarkable conserved charge $\tilde{C} = \sum_i  [q_i, p_i]$ known as the Adler-Millard charge. This charge, which has the dimensions of action, is unique to matrix dynamics [it would be trivially zero in point particle classical mechanics] and plays a central role in the emergence of quantum theory and CSL from TD. Assuming TD to be the underlying theory which we do not directly observe, one constructs its equilibrium statistical thermodynamics. As a consequence of the equipartition of the Adler-Millard charge, the canonical equilibrium average of each of the commutators in this charge satisfies the relation $[q_i,p_i]=i\hbar$: this is how the quantum commutation relations emerge without having to resort to the classical limit of quantum theory. And amazingly enough, the thermodynamic approximation to TD turns out to be the Heisenberg dynamics (equivalently Schr\"{o}dinger dynamics) of quantum theory. In this sense quantum theory is an emergent phenomenon. Next, when one considers the ever-present statistical fluctuations [Brownian motion] of the Adler-Millard charge about equilibrium, one obtains, instead of the Schr\"{o}dinger equation, a stochastic non-linear generalisation of the Schr\"{o}dinger equation, which has the same structure as the CSL model and which solves the measurement problem.  It is rather elegant that quantum theory is the thermodynamic approximation to a deeper theory, and the collapse inducing non-linear modification originates from fluctuations about equilibrium.

\medskip

$\bullet$ {\textbf{\textit {A possible resolution of the problem of time}}}: Trace Dynamics takes an external classical space-time as given. However, as we have argued above, there ought to exist an equivalent reformulation of quantum theory which does not refer to an external classical time. A generalisation of TD enables us to arrive at such a reformulation. The starting point for this generalisation is the proposal that at a more fundamental level, physical laws should be invariant under general coordinate transformations of {\it non-commuting} coordinates. Amazingly, this generalisation of Einstein's general covariance principle provides a path from classical dynamics to quantum dynamics. We first treat the simpler case of a `Minkowski' non-commuting space-time, with coordinates which are now operators $(\hat{t}, {\hat{\bf x}})$, possessing the non-commutative metric 
\begin{equation}
ds^2 = {\rm Tr} \; d\hat{s}^2 \equiv {\rm Tr}\; [c^2 d\hat{t}^2 - d\hat{\bf x}^2]
\label{nle}
\end{equation}
with the operators $(\hat t,\hat x,\hat y,\hat z)$ satisfying arbitrary commutation relations, in the spirit of trace dynamics. Matter and space-time degrees of freedom are now being treated at par. Tr stands for the ordinary matrix trace, and the proper time $s$, which we call Trace Time, plays a very important role in what follows. In this space-time, because it is non-commutative, there is no point structure or light-cone structure, nor a notion of spatial distance or temporal interval. Nonetheless, this metric is invariant under Lorentz transformations of the non-commuting operator coordinates, and one can construct a Poincar\'e invariant classical dynamics for matter degrees of freedom, which generalises special relativity, and which we call non-commutative special relativity \cite{Lochan-Singh:2011}. The canonical degrees of freedom now include, apart from the position-momentum pair $(\hat{q},\hat{p})$, also the energy-time pair $(\hat{t},\hat{E})$, and the generalised Adler-Millard charge includes also the commutator $[\hat{t}_i,\hat{E}_i]$ for every degree of freedom. Evolution is described with respect to trace time $s$. Following trace dynamics, one constructs the equilibrium statistical thermodynamics of the underlying non-commutative relativity, and obtains the canonical quantum commutation relations, along with the commutation relation $[\hat t, \hat E]=i\hbar$. It is important to note that at this emergent equilibrium level, the configuration variables, including the space-time degrees of freedom  $(\hat t,\hat x,\hat y,\hat z)$, all commute with each other, although they continue to be operators. The Generalised Quantum Dynamics [GQD] which we obtain satisfies the generalised Schr\"{o}dinger equation
\begin{equation}
i\hbar \frac{d\Psi}{ds} = H \Psi(s)
\label{gqd}
\end{equation}
Evolution is given by trace time $s$, and the configuration variables include the operator time $\hat t$ as well. This is the sought for reformulation of quantum theory which does not refer to classical time \cite{Lochan:2012}. There is no background classical space-time, with the line element still being given by (\ref{nle}).

To show that this reformulation is equivalent to quantum theory, we must first recover the classical universe with its classical space-time and classical distribution of macroscopic bodies. To do this, we bring on the scene the statistical fluctuations of the Adler-Millard charge, which result in a stochastic non-linear modification of the GQD given by (\ref{gqd}). Furthermore, now the space-time degrees of freedom  $(\hat t,\hat x,\hat y,\hat z)$ no longer commute with each other. Consider the very early universe, where there are present tiny energy-density fluctuations in the matter degrees of freedom, as well as fluctuations in the non-commutative space-time geometry. As these fluctuations grow, the non-linearity in the GQD becomes significant, resulting in a CSL like effect: matter clumps into localised classical configurations. This is {\it accompanied} by, again because of the non-linearity,  the localisation of the space-time degrees of freedom 
$(\hat t,\hat x,\hat y,\hat z)$ which can hence be mapped to c-numbers, and to a classical space-time $(t,x,y,z)$ \cite{Singh:2012}. The collapse of the wave function is responsible for the emergence of classical macroscopic bodies, and also for the emergence of classical space and time. Because the original line-element 
(\ref{nle}) is Lorentz invariant, the emergent line-element is the standard Minkowski space-time. It is rather intuitive and convincing that spontaneous collapse of the wave function, which localises macroscopic objects such as stars and galaxies on a grand scale across the entire universe, also gives rise to space: space is that which is between collapsed objects. No collapse, no space. Know collapse, know space! Not only does matter curve space-time, the spontaneous localisation of matter gives rise to space-time in the first place. 

Given this classical background, one can  show that the reformulation as GQD is equivalent to quantum theory. One can consider a micro-system either according to the generalised trace dynamics with (\ref{nle}) and arrive at (\ref{gqd}) or one can consider it according to standard trace dynamics on a classical space-time, and arrive at the Schr\"{o}dinger equation. The two are equivalent because the operators
$(\hat t,\hat x,\hat y,\hat z)$ in the GQD commute at the emergent level, and can be mapped to the ordinary
$(t,x,y,z)$. And trace time can be mapped to ordinary proper time \cite{Singh:2017}.

\medskip

$\bullet$ {\textbf{\textit {A possible resolution of the non-locality puzzle}}}: Given the machinery at hand, we can resolve the puzzle of non-locality described earlier. When a measurement is made on A to collapse its wave function, the statistical fluctuations of the Adler-Millard charge kick on, and the GQD becomes stochastic and non-linear. The space-time operators $(\hat t,\hat x,\hat y,\hat z)$ associated with the quantum system (AB) no longer commute, and {\it cannot} be mapped to the $(t,x,y,z)$ of the background space-time. Thus, the non-linear GQD, although it is invariant under Lorentz transformations of the non-commutative line element (\ref{nle}), is not invariant under the Lorentz transformations of standard Minkowski space. Thus, collapse and special relativity are incompatible with each other. GQD  can be mapped to ordinary quantum theory and is compatible with special relativity, but when the collapse inducing statistical fluctuations are added on to GQD, it is not Lorentz invariant and hence there is no relativistic theory of collapse. [In the non-relativistic limit there is an absolute time, and no need to worry about Lorentz invariance, so the CSL effect can be modelled using a random stochastic field]. We now understand why quantum theory gels so nicely with special relativity, but does not gel well when collapse is included: we are confronted with the spooky action at a distance. The non-local quantum correlation is induced by collapse, and hence can be correctly described only in the underlying non-commutative space-time (which {\it is} Lorentz invariant). In this description, since there is no notion of spatial distance between A and B, there is simply no action at a distance. Nor is there any question of an influence travelling from A to B: travelling requires distance, and here there is no distance in the first place. Micro-causality and quantum non-locality can co-exist: the former belongs to standard quantum theory, the latter belongs to the stochastic non-linear modification of standard quantum theory. This latter modification is not consistent with special relativity, but is consistent with non-commutative special relativity \cite{Banerjee:2016}.

\medskip

$\bullet$ {\textbf{\textit {What is the physical meaning of the wave function?}}}: Now we have an answer to Feynman's remark that no one understands quantum mechanics. No one understands quantum mechanics because space-time is not the best arena for describing quantum mechanics. Non-commutative space-time is. Let us revisit the double slit experiments with electrons. We have trouble, understandably, in thinking of the complex-valued wave function acting as a wave, interfering with itself in physical space: this makes no sense at all. However, the truth is that, fundamentally the wave function lives in the associated 
non-commutative line-element (\ref{nle}) of the electron. The complex-valued state of the electron belongs to the Hilbert space of the generalised (matrix) trace dynamics, which includes a non-commutative space-time. It is in no need of classical space-time for its description: the classical space-time is produced by collapsed classical objects, and is approximate. It creates problems when we try to understand quantum mechanics. But the underlying Hilbert space encompasses matter states as well as states of space-time. The troublesome distinction between space and the Hilbert space of quantum theory has been removed. But via the trace time $s$, we do have a very useful notion of time at hand. It is as if time is more fundamental than space. In the classical limit, we can attach {\it classical} observers to inertial frames, so the coordinate time $t$ plays a useful role as time. But there is no such thing as a quantum observer, so the operator time $\hat t$ is of no use to observers.

\section{Quantum theory, and space-time, are emergent phenomena}
There has been much talk lately \cite{Cowen:2015}, of space, and perhaps space-time, as emergent from quantum entanglement. But it would seem that this is hard to achieve within the confines of standard quantum theory.  
Quantum entanglement must be lost, to arrive at classical space. But to lose entanglement, we need collapse. And collapse perhaps requires us to go beyond quantum theory, and modify it.

When we go to that `beyond', we find that there is a new world out there. Both quantum theory, and space-time, are emergent phenomena. They emerge from the generalised trace dynamics. In the new world, nothing commutes with anything. Nothing is local, neither in space, nor in time. Locality and separability are approximations of our present day universe. Spontaneous collapse is omnipotent and at play, making the universe look like  it does. However, appearances can be deceptive. Deep down, if we look very carefully, everything is everywhere all the time, in a manner of speaking. Wandering is a property of the approximate universe. It is an illusion.


\newpage

\centerline{\bf TECHNICAL ENDNOTES}

\bigskip

 \noindent{ \bf Spontaneous Localisation} \cite{Ghirardi2:90,Ghirardi:86,Pearle:89,Bassi:03}

\smallskip

 \noindent Given a system of $N$ particles, with a wave function belonging to the Hilbert space, its dynamics satisfies the following properties. During its evolution, the wave function undergoes repeated spontaneous collapse at random times, mathematically described as:
\begin{equation}
{\psi_{t}({\bf x}_{1}, {\bf x}_{2}, \ldots {\bf x}_{N}) \quad
\longrightarrow \quad} 
 \frac{L_{n}({\bf x}) \psi_{t}({\bf x}_{1},
{\bf x}_{2}, \ldots {\bf x}_{N})}{\|L_{n}({\bf x}) \psi_{t}({\bf
x}_{1}, {\bf x}_{2}, \ldots {\bf x}_{N})\|}
\end{equation}

Here, $L_n({\bf x})$ is the so-called jump operator, which is defined as
\begin{equation}
L_{n}({\bf x}) =
\frac{1}{(\pi r_C^2)^{3/4}} e^{- ({\bf
q}_{n} - {\bf x})^2/2r_C^2}
\end{equation}
which localises the $n$-th particle to the spatial location ${\bf x}$ in a region of size $r_C$, with ${\bf q}_n$ being the position operator of the $n$-th particle. The probability for this jump to position 
${\bf x}$ by the $n$-th particle is given by:
\begin{equation}
p_{n}({\bf x}) \quad \equiv \quad \|L_{n}({\bf x}) \psi_{t}({\bf
x}_{1}, {\bf x}_{2}, \ldots {\bf x}_{N})\|^2
\end{equation}
The jumps are assumed to occur according to a Poisson process, with a frequency $\lambda$. Thus $\lambda$ and $r_C$ are the two new parameters of the model. Between any two  jumps the wave function undergoes normal Schr\"{o}dinger evolution. The mass density of the $n$-th particle in physical space is defined as
\begin{eqnarray}
\rho^{(n)}_{t}({\bf x}_{n}) & \equiv & m_{n} \int d^3 x_{1} \ldots
d^3 x_{n-1} d^3 x_{n+1} 
 \ldots d^3 x_{N} \, | \psi_{t}({\bf x}_{1},
{\bf x}_{2}, \ldots {\bf x}_{N}) |^2 \quad
\end{eqnarray}
These then are the axioms of the model of spontaneous collapse. It is a universal dynamics, to which quantum mechanics and classical mechanics are limiting approximations. There is no need to refer to any classical measuring apparatus, or environment, or measurement, or macroscopic world. Measurement is just a special case of spontaneous collapse. 

The beauty of spontaneous collapse is the natural manner in which the amplification mechanism comes about. In a bound system of $N$ particles, any one particle undergoing collapse causes the entire system to collapse. Thus the effective collapse rate for the system is $N\lambda$, which is an enormous amplification. If the individual particle is a nucleon, then the collapse rate can become enormously high for a macroscopic system, because $N$ is very large. Thus macroscopic objects stay effectively localised in space, and this explains their classical behaviour and solves the measurement problem.

Mathematically, spontaneous collapse can be described as a continuous process, through a stochastic 
non-linear modification of the Schr\"{o}dinger equation:
\begin{eqnarray} \label{eq:csl-massa}
d\psi_t =   \left[-\frac{i}{\hbar}Hdt \right. 
+ \frac{\sqrt{\gamma}}{m_{0}}\int d\mathbf{x} (M(\mathbf{x}) - \langle M(\mathbf{x}) \rangle_t)
dW_{t}(\mathbf{x}) \nonumber 
 -  \left. \frac{\gamma}{2m_{0}^{2}} \int d\mathbf{x}\,
(M(\mathbf{x}) - \langle M(\mathbf{x}) \rangle_t)^2 dt\right] \psi_t
\end{eqnarray}
Here, the first term describes the usual Schr\"{o}dinger evolution, with $H$ being the quantum Hamiltonian. The second and third terms are the new terms which cause dynamical collapse. The new terms are non-unitary, yet they maintain the norm-preserving nature of the evolution.   $m_0$ is a reference mass, conventionally chosen to be the mass of the nucleon. $M({\bf x})$ is the mass density operator:
\begin{eqnarray}
M(\mathbf{x})
& = & \underset{j}{\sum}m_{j}N_{j}(\mathbf{x}), \label{eq:dsfjdhz}\\
N_{j}(\mathbf{x})
& = & \int d\mathbf{y}g(\mathbf{y-x})
\psi_{j}^{\dagger}(\mathbf{y})\psi_{j}(\mathbf{y}), \qquad
\end{eqnarray}
$\psi_{j}^{\dagger}(\mathbf{y})$ and
$\psi_{j}(\mathbf{y})$ are the  creation and
annihilation operators, respectively,  for a particle $j$ at the location
$\mathbf{y}$. The smearing function $g({\bf x})$ is defined as
\begin{equation} \label{eq:nnbnm}
g(\mathbf{x}) \; = \; \frac{1}{\left(\sqrt{2\pi}r_{\text{\tiny C}}\right)^{3}}\;
e^{-\mathbf{x}^{2}/2r_{\text{\tiny C}}^{2}}
\end{equation}
The collapse inducing stochasticity in the CSL model is described by $W_{t}\left(\mathbf{x}\right)$ which is an
ensemble of independent Wiener processes, one for each point in space. The constant $\gamma$ is related to the rate parameter $\lambda$ as
\begin{equation}
\lambda_{\text{}} \; = \; \frac{\gamma}{(4\pi r_{\text{\tiny C}}^2)^{3/2}}.
\end{equation}

\bigskip

\bigskip


\medskip



\centerline{\bf REFERENCES}
\bibliography{biblioqmtstorsion}

\end{document}